\documentclass[10pt]{article}

\usepackage[cp1251]{inputenc}
\usepackage[russian, english]{babel}
\usepackage{amsmath, amssymb, amstext, amsthm,epsfig}
\usepackage[usenames]{color}
\usepackage{colortbl}

\date{}
\newcommand{\eps}{\varepsilon}

\newcommand{\N}{{\Bbb N}}

\def\Bbb#1{\mathbb #1}

\newcommand{\cnd}{\mskip 1mu|\mskip 1mu}

\renewcommand{\phi}{\varphi}
\renewcommand{\epsilon}{\varepsilon}
\renewcommand{\ge}{\geqslant}
\renewcommand{\le}{\leqslant}

\usepackage{amsthm}
\theoremstyle{plain}
\newtheorem{theorem}{Theorem}
\newtheorem{lemma}[theorem]{Lemma}

\theoremstyle{remark}
\newtheorem{remark}[theorem]{Remark}
\newtheorem{definition}[theorem]{Definition}
\newtheorem{example}[theorem]{Example}

\begin{document}

\pagestyle{plain}


\title{Algorithmic statistics: normal objects and universal models}
\author{Alexey Milovanov\\Moscow State University\\ \tt almas239@gmail.com}
\maketitle
\begin{center}
{\em {To my school teacher V.S. Shulman,
to his 70th birthday }}
\end{center}
\begin{abstract}
Kolmogorov suggested to measure 
quality of a statistical hypothesis (a model) $P$ for a data $x$
by two parameters: Kolmogorov complexity $C(P)$ of the hypothesis
and the probability $P(x)$ of $x$ with respect to $P$.
The first parameter measures how simple 
the hypothesis $P$ is and the second one how it fits.   
The paper \cite{gtv}
discovered a small class of 
models that are universal in the following sense.
Each hypothesis  $S_{ij}$ 
from that class is identified by two integer parameters 
$i,j$ and 
for every data $x$ and for each complexity level $\alpha$ there is 
a hypothesis $S_{ij}$ with $j\le i\le l(x)$ 
of complexity at most $\alpha$
that has almost the best fit 
among all hypotheses of complexity
at most $\alpha$. The hypothesis 
$S_{ij}$ is identified by $i$ and 
the leading $i-j$ bits of the binary
representation of the number 
of strings of complexity at most $i$.
On the other hand, the initial data $x$ might be completely 
irrelevant to the the number 
of strings of complexity at most $i$. Thus
$S_{ij}$ seems to have some information irrelevant to the data,
which undermines Kolmogorov's approach: the best hypotheses should not
have irrelevant information.

To restrict the class of hypotheses for a data 
$x$ to those that have only relevant information, 
the paper  \cite{nkv}
introduced a notion of a strong model for 
$x$: those are models for $x$ whose 
total conditional complexity conditional to $x$ is negligible.
An object $x$ is called normal if for each complexity
level $\alpha$ at least one its best fitting model of that
complexity is strong.

In this paper we show that there are ``many types'' of  
normal strings (Theorem \ref{VVfNS}).
Our second result states that there is a normal  
object $x$ such that all its best fitting models $S_{ij}$
are not strong for $x$. Our last result states 
that every best fit strong model for a normal object is again 
a normal object. 
\end{abstract}


\section{Introduction}

Let us recall the basic notion of algorithmic information theory 
and algorithmic statistics (see~\cite{shen15,lv,suv} for more details).
As objects, we consider strings over the binary alphabet $\{0,1\}$.
The set of all strings is denoted by $\{0,1\}^*$ and the length of
a string $x$ is denoted by $l(x)$. 
The empty string is denoted by $\Lambda$.

\subsection{Algorithmic information theory}

Let $D$ be a partial computable function mapping pairs of strings to strings.
\emph{Conditional Kolmogorov complexity} with respect to
$D$ is defined as
$$
C_D(x \cnd y)=\min\{l(p)\mid D(p,y)=x\}.
$$
In this context the function $D$ is called a \emph{description mode}
or a \emph{decompressor}. If $D(p,y)=x$
then $p$ is called a \emph{description of
$x$ conditional to $y$} or a \emph{program mapping $y$ to $x$}.

A decompressor $D$ is called \emph{universal}
if for every other decompressor $D'$ there is a string
$c$ such that $D'(p,y)=D(cp,y)$ for all $p,y$.
By Solomonoff---Kolmogorov theorem universal decompressors exist. We
pick arbitrary universal decompressor $D$ and call $C_D(x \cnd y)$ \emph{the Kolmogorov
complexity} of $x$ conditional to $y$, and denote it by $C(x \cnd y)$.
Then we define the unconditional Kolmogorov
complexity $C(x)$ of $x$ as $C(x \cnd \Lambda)$. 

By $\log n$ we denote binary logarithm. 
\emph{Symmetry of information:}
$C(x)+C(y|x)\approx C(y)+C(x|y)\approx C(x,y)$.
This equality holds with accuracy $O(\log(C(x)+C(y))$
and is due to Kolmogorov and Levin.





\subsection{Algorithmic statistics: basic notions}

Algorithmic statistics studies  explanations of observed data that are suitable in the algorithmic sense: an explanation should be simple and capture all the algorithmically discoverable regularities in the data. The data is encoded, say, by a binary string $x$. In this paper we consider explanations (statistical hypotheses) of the form ``$x$ was drawn at random from a finite set $A$ with uniform distribution''.

Kolmogorov suggested in a talk~\cite{kolm} in 1974
to  measure the quality of an explanation
$A\ni x$ by two parameters:  Kolmogorov
complexity $C(A)$\footnote{Kolmogorov complexity of $A$ is defined as follows. We fix any computable bijection $A \to [A]$ from the family of all finite sets to the set of binary strings, called \emph{encoding}. Then we define $C(A)$ as the complexity $C([A])$ of the code $[A]$ of $A$.
In a similar  we define Kolmogorov complexity of other finite objects.} of $A$
and the log-cardinality $\log_2|A|$  of $A$.
The smaller $C(A)$ is the simpler the explanation is.
The  log-cardinality measures the \emph{fit} of $A$---the lower is
$|A|$ the more $A$ fits as an explanation for any of its elements.
For each complexity level $m$ any model 
$A$ for $x$ with smallest  $\log|A|$ among models
of complexity at most $m$ for $x$ is called a \emph{best
fit hypothesis for $x$}. The 
trade off between  $C(A)$ and $\log|A|$
is represented by 
the \emph{profile} of $x$.

\begin{definition}
The  \emph{profile} of a string $x$ is the set $P_x$
consisting of all pairs $(m, l)$ of natural numbers such 
that there exists a finite set $A \ni x$ 
with $C(A) \le m$ and $\log_2|A| \le l$.
\end{definition}

\begin{figure}[h]
\begin{center}
\includegraphics[scale=1]{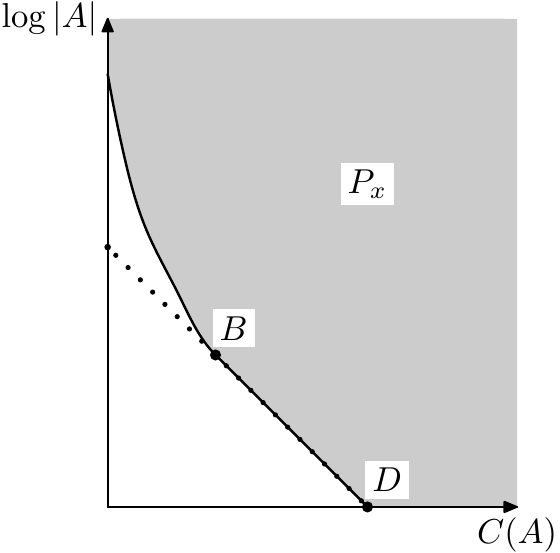}
\end{center}
\caption{The profile $P_x$ of a string $x$.}
\label{f1}
\end{figure}

Both parameters $C(A)$ and $\log|A|$ cannot be very small simultaneously
unless the string $x$ has very small Kolmogorov complexity. Indeed,
$C(A)+\log_2|A|\gtrsim C(x)$, since $x$
can be specified by $A$ and its index in $A$.
A model $A \ni x$ is called \emph{sufficient or optimal} if 
$C(A) + \log|A| \approx C(x)$. 
The value 
$$
\delta(x \cnd A) = C(A) + \log|A| - C(x)
$$
is called the \emph{optimality deficiency} of $A$ as 
a model for $x$.
On Fig. \ref{f1} parameters of sufficient statistics lie on the segment $BD$.
A sufficient statistic that has the minimal complexity is called 
\emph{minimal} (MSS), its parameters are represented 
by the point $B$ on  Fig. 1.

\begin{example}
Consider a string $x \in \{0, 1\}^{2n}$ such that leading $n$ bits of $x$  
are zeros, and the remaining bits are random, 
i. e. $C(x) \approx n$. 
Consider the model $A$ for $x$ 
that consists of all strings from $\{0, 1\}^{2n}$ 
that have $n$ leading zeros. 
Then 
$C(A) + \log|A|=\log n+O(1)+n\approx C(x)$, hence $A$ is a 
sufficient statistic for $x$. As the complexity of $A$
is negligible, $A$ is a minimal sufficient statistic for $x$.
\end{example}

The string from this example has a sufficient statistic 
of negligible complexity. Such strings  are called \emph{stochastic}. 
Are there strings that have no sufficient statistics of negligible complexity?
The positive answer to this question
was obtained in \cite{shen}. Such strings are called \emph{non-stochastic}. 
Moreover, under some natural constraints for every 
set $P$ there is a string whose profile is close to $P$.
The constraints are listed in the following theorem:

\begin{theorem}
\label{th1}
Let $x$ be a string of length $n$ and complexity $k$.
Then $P_x$ has the following properties:

1) $(k+O(\log n), 0) \in P_x$.

2) $(O(\log n), n) \in P_x$.

3) if $(a,b + c) \in P_x$ then $(a + b+O(\log n), c) \in P_x$.

4) if $(a, b) \in P_x$ then $a + b > k - O(\log n)$.
\end{theorem}

In other words, with logarithmic accuracy,
the boundary of $P_x$ 
contains a point $(0, a)$ with $a \le l(x)$, 
contains the point $(C(x), 0)$,
decreases with the slope at least $-1$
and lies above the line $C(A)+\log|A|=C(x)$.
Conversely, given a curve with these property
that has low complexity one can find a string $x$ of length $n$
and complexity about $k$ such that the boundary of $P_x$ is close to
that curve:
\begin{theorem}[\cite{vv}]
\label{th2}
Assume that we are  given $k,n$ and an upward closed
set $P$ of pairs of natural numbers
such that $(0,n),(k,0)\in P$,
$(a,b+c)\in P\Rightarrow (a+c,b)\in P$
and $(a,b)\in P\Rightarrow a+b\ge k$.
Then there is a string $x$
of length $n$ and complexity $k+O(\log n)$ whose profile is 
$C(P)+O(\log n)$-close to $P$.
(We call subsets of
$\N^2$ \emph{$\eps$-close} if each of them is in
the $\eps$-neighborhood of the other.)
By $C(P)$ we denote the Kolmogorov complexity of the boundary 
of $P$, which is a finite object.
\end{theorem}

\subsection{Models of restricted type}
\label{restrictedmodels}
It turns out that Theorems~\ref{th1} and~\ref{th2}
remain valid (with smaller accuracy) even if we restrict
the class of models under consideration to models from
a class $\mathcal A$ provided the class $\mathcal A$ 
has the following properties.

(1) The family $\mathcal{A}$
is enumerable. This means that there exists an algorithm
that prints elements of
$\mathcal{A}$  as lists of strings, 
with some separators (saying where one element of
$\mathcal{A}$ ends and another one begins).

(2) For every $n$ the class $\mathcal{A}$ contains the set $\{0,1\}^n$.

(3) The exists some polynomial
$p$ with the following property: for every $A \in \mathcal{A}$, 
for every natural $n$ and for every natural $c < |A|$ 
the set of all $n$-bit strings in $A$ can be covered by 
at most $p(n)\cdot |A|/c$ sets of cardinality at most $c$ from $\mathcal{A}$.

Any family of finite sets  of strings
that satisfies these three conditions is called \emph{acceptable}. 

Let us define the \emph{profile of $x$ with respect to $\mathcal A$}:
$$
P_x^{\mathcal{A}}=\{(a, b) \mid \exists A  \ni x: A \in \mathcal{A}, C(A) \le a, \log|A| \le b\}.
$$
Obviously $P_x^{\mathcal{A}} \subseteq P_x$.
Let us fix any acceptable class $\mathcal A$ of models.

\begin{theorem}[\cite{rdf}]
\label{th1a}
Let $x$ be a string of length $n$ and complexity $k$.
Then $P_x^{\mathcal{A}}$ has the following properties:

1) $(k+O(\log n), 0) \in P_x^{\mathcal{A}}$.

2) $(O(\log n), n) \in P_x^{\mathcal{A}}$.

3) if $(a,b + c) \in P_x$ then $(a + b+O(\log n), c) \in P_x^{\mathcal{A}}$.

4) if $(a, b) \in P_x^{\mathcal{A}}$ then $a + b > k - O(\log n)$.
\end{theorem}

\begin{theorem}[\cite{rdf}]
\label{th2a}
Assume that we are  given $k,n$ and an upward closed
set $P$ of pairs of natural numbers
such that $(0,n),(k,0)\in P$,
$(a,b+c)\in P\Rightarrow (a+c,b)\in P$
and $(a,b)\in P\Rightarrow a+b\ge k$.
Then there is a string $x$
of length $n$ and complexity $k+O(\log n)$ such 
that \emph{both sets}  
$P_x^{\mathcal{A}}$ and $P_x$ are 
$C(P)+O(\sqrt{ n\log n})$-close to $P$.
\end{theorem}

\begin{remark}
Originally, the conclusion of Theorem~\ref{th2a}
stated only that the set $P_x^{\mathcal{A}}$ is close to the given set $P$.
However, as observed in \cite{suv}, 
the proof from~\cite{rdf} shows also that 
$P_x$ is close to $P$.
\end{remark}

\subsection{Universal models}
\label{um}

Assume that $A$ is sufficient statistic $A$ for $x$. Then $A$
provides
a two-part code $y=($the shortest description of $A$, the index of $x$ in $A$) 
for $x$ whose total length is close to the complexity of $x$.
The symmetry of information implies that 
$C(y \cnd x)\approx C(y)+C(x \cnd y)-C(x)$. Obviously,
the term $C(x \cnd y)$ here is negligible and 
$C(y)$ is at most its total length, which by assumption is
close $C(x)$. 
Thus $C(y|x)\approx 0$, that is, $x$ and $y$ have almost the same information.
That is, the two-part code $y$ for $x$
splits the information from $x$ in two parts:
the shortest description of $A$, the index of $x$ in $A$. 
The second part of this two-part code 
is incompressible (random) conditional to the first 
part (as otherwise, the complexity of $x$
would be smaller the the total length of $y$).
Thus  the second part of this
two-part code 
can be considered as accidental information  (noise) in the data $x$. 
In a sense every sufficient statistic $A$ identifies 
about $C(x)-C(A)$ bits of accidental information in $x$.
And thus any minimal sufficient statistic for $x$ 
extracts almost all useful information from $x$.

However, it turns out that this viewpoint is inconsistent
with the existence of universal models, discovered in~\cite{gtv}.
Let $L_m$ denote the list of strings of complexity at most $m$.
Let $p$ be an algorithm that
enumerates all strings of $L_m$ in some order. 
Notice that there is such algorithm of complexity 
$O(\log m)$.
Denote by $\Omega_m$ the cardinality of $L_m$.   
Consider its binary representation, i. e., the sum:
$$\Omega_m = 2^{s_1}+2^{s_2}+...+ 2^{s_t} \text{, where }s_1 > s_2>...> s_t.$$
According to this decomposition and $p$, we split $L_m$ into groups:
first $2^{s_1}$ elements, next $2^{s_2}$ elements, etc.  Let us denote by $S_{m,s}^p$ the group of size $2^s$ from the partition. Notice that $S_{m,s}^p$ is defined only for $s$ that correspond to ones in the binary
representation of $\Omega_m$, so $m \ge s$.

If $x$ is a string of complexity at most $m$, it belongs to some group $S_{m,s}^p$
and this group can be considered as a model for $x$. We may consider different
values of $m$ (starting from $C(x)$). In this way we get different models $S^p_{m,s}$
for the same $x$. The complexity of $S^p_m$ is $m -  s + O(\log m+C(p))$. 
Indeed, chop
$L_m$ into portions of size $2^s$ each, then $S^p_{m,s}$ is the last full portion and can
be identified by $m$, $s$ and the number of full portions, which is less than
$\Omega_m / 2^s < 2^{m-s+1}$.
Thus if $m$ is close to $C(x)$ and $C(p)$ is small 
then $S^p_{m,s}$ is a sufficient statistic for $x$.
More specifically $C(S^p_{m,s})+\log|S^p_{m,s}|=C(S^p_{m,s})+s=m+O(\log m+C(p))$.

For every $m$ there is an algorithm $p$ of complexity $O(\log m)$ 
that enumerates all strings of complexity at most $m$.
We will fix for every $m$ any such algorithm $p_m$ 
and denote $S^{p_m}_{m,s}$ by $S_{m,s}$.

The models $S_{m,s}$ were introduced in~\cite{gtv}.
The models $S^p_{m,s}$ are universal in the following sense:
\begin{theorem}[\cite{vv}]\footnote{This 
theorem was proved in~\cite[Theorem VIII.4]{vv}
with accuracy $O(\max\{\log C(y)\mid y\in A\}+C(p))$
instead of $O(\log n)$. Applying \cite[Theorem VIII.4]{vv}
to $A'=\{y\in A\mid l(y)=n\}$ we obtain
the theorem in the present form.}\label{l8}
Let $A$ be any finite set of strings containing a string 
$x$ of length $n$. 
Then for every $p$ there are $s\le m \le n + O(1)$ such that\\
1) $x \in S^p_{m,s}$,\\
2) $C(S^p_{ m,s} \cnd A) = O(\log n+C(p))$ (and hence 
$C(S^p_{m,s})\le C(A)+O(\log n+C(p))$),\\
3) $\delta(x \cnd S^p_{m,s}) \le \delta(x \cnd A) + O(\log n+C(p))$.
\end{theorem}

In turns out 
that the model $S^p_{m,s}$ has the same information 
as the the number $\Omega_{m-s}$:
\begin{lemma}[\cite{vv}]\label{l7}
For every $a \le b$ and for every $s \le m$:

1) $C(\Omega_a \cnd \Omega_b) = O(\log b)$.

2) $C(\Omega_{m-s} \cnd S^p_{m,s}) = O(\log m+C(p))$ and 
$C(S^p_{m,s} \cnd \Omega_{m-s}) = O(\log m+C(p))$.

3) $C(\Omega_a) = a + O(\log a)$.
\end{lemma}

By Theorem~\ref{l8} for every data $x$ there is a minimal sufficient
statistic for $x$ of the form $S_{m,s}$. Indeed, let $A$ be any 
minimal sufficient statistic for $x$ and let  $S_{m,s}$ be any model
for $x$ that exists by Theorem~\ref{l8} for this $A$.
Then by item 3 the statistic $S_{m,s}$ is sufficient as well
and by item 2 its complexity is also close to minimum.
Moreover, since $C(S_{m,s} \cnd A)$ is negligible
and $C(S_{m,s})\approx C(A)$, by symmetry of information
$C(A \cnd S_{m,s})$ is negligible as well. Thus $A$ has the same
information as $S_{m,s})$, which has the same information
as $\Omega_{m-s}$ (Lemma~\ref{l7}(2)). 
Thus if we agree that every minimal sufficient statistic
extracts all useful information from the data,
we must agree also that that information is the same as
the information in the number of strings of complexity
at most $i$ for some $i$.

\subsection{Total conditional complexity and strong models}
The paper~\cite{nkv} suggests the following 
explanation to this paradox.
Although conditional complexities
$C(S_{m,s} \cnd A)$ and $C(S_{m,s} \cnd x)$ are small,
the short programs that map 
$A$ and $x$, respectively, to $S_{m,s}$ work
in a huge time. A priori their work time is not bounded by 
any total computable function of their input.
Thus it may happen that 
practically we are not able to find $S_{m,s}$ (and also $\Omega_{m-s}$)
from a MSS $A$ for $x$ or from $x$ itself.

Let us consider now programs
whose work time is bounded by a total 
computable function for the input.
We get the notion of \emph{total conditional complexity} $CT(y\cnd x)$,
which is the length of the shortest \emph{total} program
that maps $x$ to $y$.  
Total conditional complexity can be much greater than plain one,
see for example \cite{shen12}. 
Intuitively, good sufficient statistics $A$ for $x$ 
must have not only negligible conditional complexity
$C(A \cnd x)$ (which follows from definition of a sufficient
statistic) but also negligible \emph{total}
conditional complexity $CT(A \cnd x)$.
The paper~\cite{nkv} calls such models $A$ \emph{strong models for $x$}.

Is it true that for some $x$ there is no \textbf{strong} MSS
$S_{m,s}$ for $x$? The positive answer to this question was obtained
in~\cite{nkv}: there are strings $x$ whose all minimal 
sufficient statistics are not strong for $x$.
Such strings are called \emph{strange}.
In particular, if $S_{m,s}$ is a MSS for strange string
$x$ then  $CT(S_{m,s} \cnd x)$ 
is large.  However, a strange string 
has no strong MSS at all. An interesting
question is whether there are strings 
$x$ that do have strong  MSS but  
have no strong MSS of the form $S_{m,s}$? 
This question was left open in~\cite{nkv}.
In this paper we answer this question in positive.
Moreover, we show that there is a ``normal'' string
$x$ that has no strong MSS of the form $S_{m,s}$ (Theorem~\ref{nswnsmss}).
A string $x$ is called  \emph{normal} if 
for every complexity level  
$i$ there is a best fitting model $A$ for $x$
of complexity at most $i$ (whose parameters
thus lie on the border of the set $P_x$) that is strong.
In particular, every normal string has a strong MSS. 

Our second result 
answers yet another question asked in~\cite{nkv}. 
Assume that $A$ is a strong  
MSS for a normal string $x$. Is it true that the code $[A]$ of $A$ 
is a normal string itself? Our Theorem \ref{hereditary-theorem} states 
that this is indeed the case. Notice that by a result
of~\cite{nkv} the profile  $P_{[A]}$ of $[A]$ can be obtained from
$x$ by putting the origin in the point corresponding to parameters 
of $A$ (i.e. in the point $B$ on Fig.~\ref{f1}). 

Our last result (which comes first in the following exposition) 
states that there are normal strings
with any given profile, under the same restrictions as in Theorem~\ref{th1}
(Theorem~\ref{VVfNS} in Section \ref{fixcurve}).


\section{Normal strings with a given profile}
\label{fixcurve}
In this section we prove an analogue of Theorem~\ref{th2} 
for normal strings. We start with a 
rigorous 
definitions of strong models and normal  strings. 
 
A set $A \ni x$ is called \emph{$\epsilon$-strong statistic (model)
for a string $x$} if $CT(A \cnd x) < \epsilon$. 
To represent the trade off between size and complexity of 
$\eps$-strong models for $x$ 
consider the \emph{$\eps$-strong profile of $x$}: 
$$ 
P_x^\epsilon = \{(a,b) \mid \exists A \ni x: CT(A\cnd x)\le \epsilon, 
C(A) \le a, \log|A| \le b \}.
$$

It is not hard to see that the set
$P_x^\epsilon $ satisfies the item (3) from Theorem~\ref{th1}:
\begin{quote}
for all $x \in \{0,1\}^n$ if
$(a, b + c) \in P_x^{\epsilon}$ then 
$(a + b + O(\log n), c) \in P_x^{\epsilon + O(\log n)}$.
\end{quote}

It follows from the definition that 
$P_x^{\epsilon}\subset P_x$ for all $x,\eps$.
Informally a string is called normal if for a negligible 
$\eps$ we have $P_x \approx P_x^{\epsilon}$. 
Formally, for integer parameters $\eps,\delta$ we say 
that a string $x$ is \emph{$\epsilon, \delta$-normal}  
if $(a,b) \in P_x$ implies $(a + \delta, b + \delta) \in P_x^{\epsilon}$
for all $a,b$.
The smaller  $\eps,\delta$ are the stronger
is the property of $\eps,\delta$-normality. 
The main result of this section shows that for some 
$\eps,\delta=o(n)$ for every set $P$ satisfying 
the assumptions of Theorem~\ref{th1} there is an
$\eps,\delta$-normal string of length $n$ with
$P_x\approx P$: 

\begin{theorem}
\label{VVfNS}
Assume that we are  given an upward closed
set $P$ of pairs of natural numbers
satisfying assumptions of Theorem~\ref{th2}.
Then there is an $O(\log n),O(\sqrt{n\log n})$-normal string $x$
of length $n$ and complexity $k+O(\log n)$ whose profile $P_x$ is 
$C(P)+O(\sqrt{n\log n})$-close to $P$.
\end{theorem}
\begin{proof}
We will derive this theorem from Theorem~\ref{th2a}.
To this end consider the following family $\mathcal{B}$ of sets.
A set  $B$ is in this family if it has the form
$$
B=\{uv\mid v\in\{0,1\}^m\},
$$ 
where $u$ is an arbitrary binary string and $m$ is an arbitrary 
natural number.
Obviously, the family $\mathcal{B}$ is  acceptable,
that is, it satisfies the properties (1)--(3)
from Section~\ref{restrictedmodels}.

Note that for every $x$ and for every $A \ni x$ from $\mathcal{B}$ 
the total complexity of $A$ given $x$ is $O(\log n)$. 
So $P_x^{\mathcal{B}} \subseteq P_x^{O(\log n)}$.  
By Theorem~\ref{th2a} there is a string $x$  such that $P_x$ and 
$P_x^{\mathcal{B}}$ are 
$C(P)+O(\sqrt{n\log n})$-close to $P$. 
Since $P_x^{\mathcal{B}} \subseteq P_x^{O(\log n)} \subseteq P_x$ we 
conclude that $x$ is $O(\log n),O(\sqrt{n\log n})$-normal.
\end{proof}

The proof of Theorem~\ref{VVfNS} 
is based on a technically difficult Theorem~\ref{th2a}.
However, for some sets $P$ it can be shown directly with
a better accuracy of $O(\log n)$ in place of $O(\sqrt{n \log n}))$.
For instance, this happens for the smallest set $P$,
satisfying the assumptions of Theorem~\ref{th2a},
namely for the set 
$$
P=\{(m,l)\mid m\ge k, \text{ or } m+l\ge n\}.
$$
Strings with such profile are called 
``antistochastic''. 
\begin{definition}
A string $x$ of length $n$ and complexity
$k$ is called \emph{$\eps$-antistochastic}
if for all $(m,l)\in P_x$ either
$m>k-\eps$, or $m+l>n-\eps$. 
\end{definition}
\begin{figure}[h]
\begin{center}
\includegraphics[scale=1]{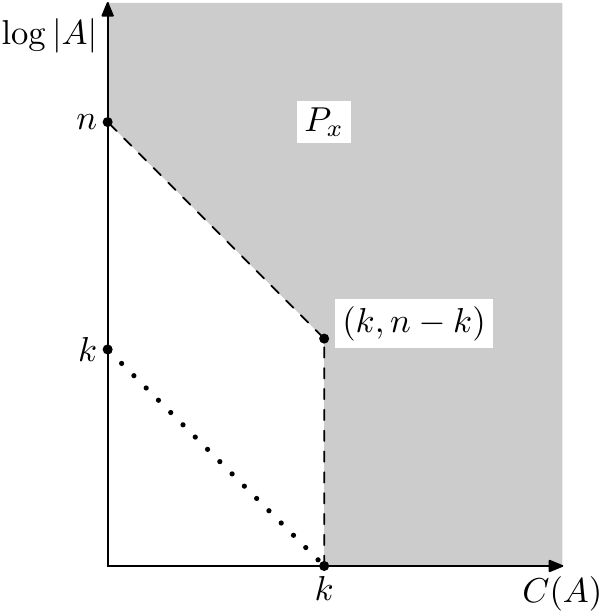}
\end{center}
\caption{The profile of an $\eps$-antistochastic string $x$ for a 
small $\eps$. }
\label{f2}
\end{figure}

We will need later the fact that for every $n$ there is
an $O(\log n)$-antistochastic string $x$ of
length $n$ and that such strings are normal:
\begin{lemma}[Proved in Appendix]\label{pan}
For all $n$ and all $k\le n$
there is an $O(\log n)$-antistochastic string $x$ of length $n$ and complexity
$k+O(\log n)$.
Any such string $x$ is $O(\log n)$,$O(\log n)$-normal.
\end{lemma}

\section{Normal strings without
universal MSS}
\label{without}

Our main result of this section is Theorem~\ref{nswnsmss} 
which states that there is a normal string $x$  
such that no set $S_{m,l}$ is not a strong MSS for $x$. 

\begin{figure}[h]
\begin{center}
\includegraphics[scale=1]{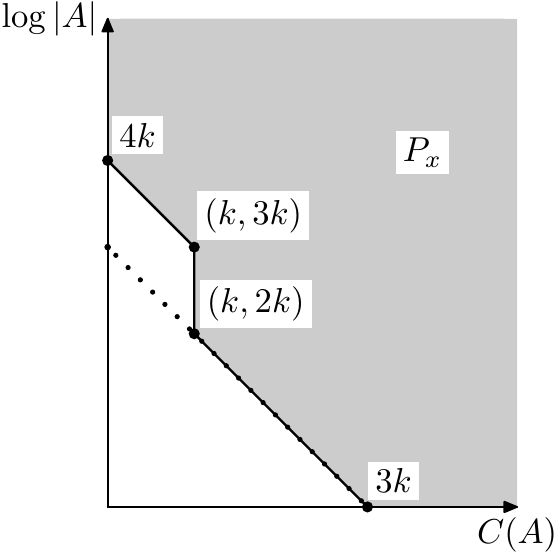}
\end{center}
\caption{The profile $P_x$ of a string $x$ from Theorem \ref{nswnsmss}.}
\label{f4}
\end{figure}

\begin{theorem}
\label{nswnsmss}
For all large enough $k$ there exist an $O(\log k)$-normal string $x$ 
of complexity $3k+O(\log k)$ and length $4k$ such that:

1) The profile $P_x$ of $x$ is $O(\log k)$-close
to the gray set on Fig.~\ref{f4}.

2) The string $x$ has a strong MSS. 
More specifically, there is an $O(\log k)$-strong model $A$ for $x$ 
with complexity $k+O(\log k)$ and log-cardinality $2k$.

3) For all simple $q$ and all $m,l$ the set
$S^q_{m,l}$ cannot be a strong sufficient statistic for $x$.
More specifically, 
for every $\epsilon$-strong $\eps$-sufficient 
model $S^q_{m,l}$ for $x$
of complexity at most $k+\delta$
we have $O(\epsilon + \delta + C(q)) \ge k - O(\log k)$.
\end{theorem}

In the proof of this theorem
we will need a rigorous definition of MSS and a related result from \cite{nkv}. 

\begin{definition}
A set $A$ is called a $\delta, \epsilon, D$-minimal sufficient statistic (MSS) for $x$ if $A$ is 
an $\eps$-sufficient statistic for $x$ and there is no model $B$ for $x$ with $C(B)<C(A)-\delta$
and $C(B)+\log|B|-C(x)<\eps+D\log C(x)$.
\end{definition}

The next theorem 
states that for every strong MSS $B$ and for every sufficient statistic
$A$ for $x$ the total conditional complexity $CT(B \cnd A)$ is negligible.

\begin{theorem}[\cite{nkv}, Theorem 13]
\label{MSSandSS}
For some constant $D$ if
$B$ is $\eps$-strong $\delta, \epsilon, D$-minimal 
sufficient statistic for $x$ 
and $A$ is an $\eps$-sufficient statistic for $x$ then
$CT(B \cnd A)=O(\eps+\delta+\log C(x))$.
\end{theorem}

Let us fix a constant $D$ satisfying Theorem \ref{MSSandSS} 
and call a model $\delta, \epsilon$-MSS if it is 
$\delta, \epsilon, D$-MSS. 
Such models have the following property.
\begin{theorem}[\cite{nkv}, Theorem 14]
\label{thprg}
Let $x$ be a string of length $n$ and $A$ be an $\epsilon$-strong 
$\eps$-sufficient 
statistic for $x$.
Then for all $b \ge \log|A|$ we have 
$$
(a, b) \in P_x \Leftrightarrow (a + O(\epsilon + \log n), b - \log |A| + O(\epsilon + \log n)) \in P_{[A]} 
$$
and for $b\le \log|A|$ we have 
$(a, b) \in P_x \Leftrightarrow a+b\ge C(x)-O(\log n)$.
\end{theorem}

\begin{proof}[The proof of Theorem~\ref{nswnsmss}]
Define $x$ as the concatenation of strings $y$ and $z$, where $y$ is an 
$O(\log k)$-antistochastic string 
of complexity $k$ and length $2k$ (existing by Lemma~\ref{pan})
and $z$ is a string of 
length $2k$ such that $C(z \cnd y) \ge 2k - O(\log k)$ (and 
hence $C(x) = 3k + O(\log k)$).  
Consider the following set $A=\{yz'\mid l(z')=2k\}$. From the shape of $P_x$
it is clear that $A$ is an  
$O(\log k),O(\log k)$-MSS  for $x$. 
Also it is clear that $A$ is  an $O(\log k)$-strong model for $x$. 
So, by Theorem \ref{thprg} the profile of $x$  is $O(\log k)$-close to the gray set on Fig. \ref{f4}. From normality of $y$ (Lemma \ref{pan}) 
it is not difficult to see that $x$ 
is $O(\log k)$-normal.

Let $S^q_{m,l}$ be an $\epsilon$-strong $\epsilon$-sufficient 
model for $x$ of complexity at most $k+\delta$. 
We claim that 
$S^q_{m,l}$ is an $\eps,\delta+O(\log k)$-MSS for $x$. 
In other words, 
$C(B)\le C(S^q_{m,l})-\delta - O(\log k)$ implies 
$C(B)+\log|B|>C(x)+\eps +D \log k$ 
where  $D$ is the constant from Theorem~\ref{MSSandSS}.

The assumption $C(B)\le C(S^q_{m,l})-\delta - O(\log k)$ 
and the assumed upper bound for $C(S^q_{m,l})$ imply
that $C(B)\le k- O(\log k)$. 
From the shape of $P_x$ it follows that 
$C(B)+\log|B|\ge C(x)+k-O(\log k)$.
Notice that if $\eps$ is close to $k$  the conclusion 
of the theorem is straightforward. 
Otherwise, the last inequality implies 
$C(B)+\log|B|>C(x)+ \eps +D \log k$.

By Theorem \ref{MSSandSS} we get
$CT(S_{m,l}^q \cnd A) = O(\eps + \delta +\log k)$ and 
thus 
$CT(s_0 \cnd y) = O(\eps + \delta +\log k)$, where $s_0$ is the 
lexicographic least element in $S_{m,l}^q$.
Denote by $p$ a total program of length  $O(\eps + \delta +\log k)$ 
that transforms $y$ to  $s_0$. 
Consider the following set $B:= \{p(y') \mid l(y')=2k \}$. We claim that if $\epsilon $ and $\delta$ are not very big, then the complexity of any element from $B$ is not greater than $m$. Indeed, if $\epsilon + \delta \le d k$ for 
a small constant $d$, then $l(p) < k - O(\log k)$ and hence every element from $B$ has complexity  
at most $C(B) + \log|B| + O(\log k) \le 3k-O(\log k) \le m$.
The last inequality holds because $S_{m,l}^q$ is a model for $x$
and hence $m\ge C(x)=3k+O(\log k)$.

Let us run the program
$q$ until it prints all elements from $B$. 
Since $s_0 \in B$, there are at most $2^l$ elements of complexity $m$ 
that we have not been printed yet. So, we can find the list of all strings of 
complexity at most $m$ from $B$, $q$ and some extra
$l$ bits. 
Since this list has complexity at least $m - O(\log m)$ 
(as from this list and $m$ we can compute a string of complexity more than $m$),
we get
$O(C(B) + C(q)) + l \ge m - O(\log m)$. 

Recall that the $C(S_{m,l}^q)+\log|S_{m,l}^q|$ is equal to $m+O(\log m+C(q))$ and
is at most $C(x)+\eps$. Hence $m\le 4k$ unless 
$\eps>k+O(\log k+C(q))$. Therefore the term $O(\log m)$
in the last inequality can be re-written as $O(\log k)$.

Recall that the complexity of $S_{m,l}^q$ is $m-l+O(\log m +C(q))$.
From the shape of $P_x$ it follows that 
$C(S_{m,l}^q)\ge k-O(\log k)$ or 
$ C(S_{m,l}^q)+\log|S_{m,l}^q|\ge C(x)+k-O(\log k)$.
In the latter case $\eps \ge k-O(\log k)$ and we are done.
In the former case $m-l\ge k-O(\log k +C(q))$ and hence
$O(C(B) + C(q))\ge  k-O(\log k +C(q))$.
 \end{proof}

\section{Hereditary of normality}
\label{hereditary}
In this section we prove that every strong MSS for a normal 
string is itself normal. 
Recall that a string $x$ is called $\epsilon,\delta$-normal if 
for every model $B$ for $x$ there is 
a model $A$ for $x$ with $CT(A|x)\le\eps$ and   
$C(A)\le C(B)+ \delta$, $\log|A|\le\log|B|+ \delta$.

\begin{theorem}
\label{hereditary-theorem}
There is a constant $D$ such that the following holds.
Assume that $A$ is an $\epsilon$-strong  
$\delta, \epsilon,D$-MSS for an 
$\epsilon,\eps$-normal string $x$ of length $n$. 
Assume that $\epsilon \le \sqrt{n}/2$. 
Then the code $[A]$ of $A$ 
is $O((\epsilon + \delta + \log n)\cdot \sqrt{n})$-normal.
\end{theorem}

The rest of this section is the proof of this theorem.
We start with the following lemma, which  
a simple corollary of Theorem~\ref{l8}
and Lemma~\ref{l7}. For the sake of completeness we prove it in Appendix.
\begin{lemma}
\label{MSSOmega}
For all large enough $D$ the following holds:
if $A$ is a $\delta, \epsilon, D$-MSS for $x \in \{0,1\}^n$  
then $C(\Omega_{C(A)} \cnd A) = O(\delta + \log n).$
\end{lemma} 

We fix a constant $D$ satisfying 
Lemma~\ref{MSSOmega} and call a model $\delta, \epsilon$-MSS if it 
$\delta, \epsilon, D$-MSS. This $D$ is the constant satisfying
Theorem~\ref{hereditary-theorem}

A family of sets $\mathcal{A}$ is called \emph{partition} if 
for every $A_1, A_2 \in \mathcal{A}$ 
we have $A_1 \cap A_2 \not= \varnothing \Rightarrow A_1 = A_2$. 
Note that for a finite partition we can define its complexity.  
The next lemma states that every strong statistic $A$  
can be transformed to a strong statistic  $A_1$ 
such that $A_1$ belongs to some simple partition.
 
\begin{lemma}
\label{part}
Let $A$ be an $\epsilon$-strong statistic for $x \in \{0,1\}^n$. Then there is a set $A_1$ and a partition $\mathcal{A}$ 
of complexity at most  $\epsilon + O(\log n)$ such that:

1) $A_1$ is $\epsilon + O(\log n)$-strong statistic for $x$.

2) $CT(A \cnd A_1) < \epsilon + O(\log n)$ and $CT(A_1 \cnd A) < \epsilon + O(\log n)$.

3) $|A_1| \le  |A|$.

4) $A_1 \in \mathcal{A}$.
\end{lemma}
\begin{proof}
Assume that $A$ is an $\epsilon$-strong statistic for $x$.
Then there is a total program $p$ such that 
$p(x) = A$ and $l(p) \le \epsilon$.  

We will use the same construction as in Remark 1 in \cite{nkv}. 
For every set $B$ denote by $B'$ the following set: 
$\{x' \in B \mid p(x') = B,\ x' \in \{0,1\}^n \}$. 
Notice that $CT(A' \cnd A)$, $CT(A \cnd A')$ and 
$CT(A' \cnd x )$ are less than 
$l(p) + O(\log n) = \epsilon + O(\log n)$ and $|A'|\le |A|$.

For any $x_1, x_2 \in \{0,1\}^n$ with 
$p(x_1) \ne p(x_2)$ we have $p(x_1)' \cap  p(x_2)' = \varnothing$. 
Hence $\mathcal{A}:= \{p(x)'\cnd x \in \{0,1\}^n \}$ 
is a partition of complexity at most  $\epsilon + O(\log n)$. 
\end{proof}

By Theorem~\ref{l8} and Lemma~\ref{l7} 
for every $A \ni x$ there is a $B \ni x$ 
such that $B$ is informational equivalent $\Omega_{C(B)}$ and parameters of
$B$ has are not worse than those of $A$.
We will need a similar result for normal strings and for strong models. 
\begin{lemma}
\label{lch}
Let $x$ be an $\epsilon, \alpha$-normal string with length $n$ such that $\epsilon \le n$, $\alpha < \sqrt{n}/2$ . Let $A$ be an $\epsilon$-strong  statistic for $x$.  Then there is a set $H$ such that:

1) $H$ is an $\epsilon$-strong statistic for $x$.

2) $\delta(x \cnd H) \le \delta(x \cnd A) + O((\alpha + \log n)\cdot \sqrt{n})$ and $C(H) \le C(A)$.

3) $C(H \cnd \Omega_{C(H)}) = O(\sqrt{n})$. 
\end{lemma} 

\begin{proof}[Sketch of proof (the detailed proof is deferred to Appendix)]
Consider the sequence $A_1, B_1, A_2, B_2, \dots $ of statistics for $x$
defined as follows.
Let $A_1:=A$ and let $B_i$ be an improvement of $A_i$ 
such that $B_i$ is informational equivalent to $\Omega_{C(B_i)}$, which
exists by Theorem~\ref{l8}.
Let $A_{i+1}$ be a strong statistic for $x$ that has a similar parameters as 
$B_i$, which exists because $x$ is normal. (See Fig. \ref{f5}.)
\begin{figure}[h]
\begin{center}
\includegraphics[scale=1]{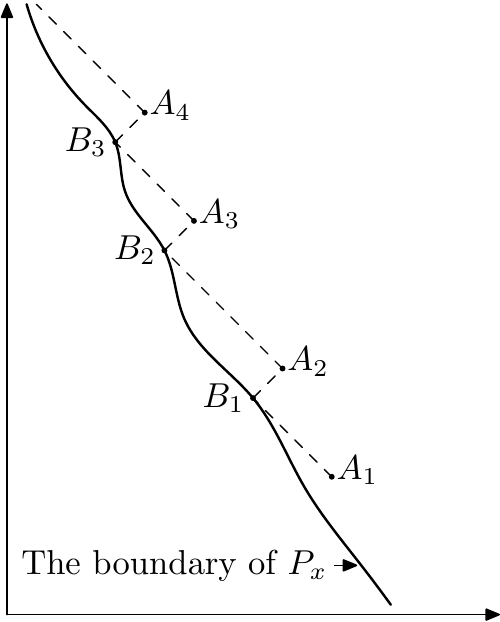}
\end{center}
\caption{Parameters of statistics $A_i$ and $B_i$}
\label{f5}
\end{figure}

Denote by $N$ the minimal integer such that $C(A_N) - C(B_N) \le \sqrt{n}$. 
For $i<N$ the complexity of $B_i$ is more than $\sqrt n$ less 
that that of $A_i$. On the other hand, the complexity  of 
$A_{i+1}$  is at most $\alpha<\sqrt n/2$
larger than that of $B_i$.
Hence $N = O(\sqrt{n})$. 
Let $H:=A_N$. By definition $A_N$ (and $H$) is strong. From $N = O(\sqrt{n})$ it follows that the second condition is satisfied. From $C(A_N) - C(B_N) \le \sqrt{n}$ and definition of $B_N$ it is follows that the third condition  is satisfied too (use symmetry of information).  
\end{proof}

\begin{proof}[Sketch of proof of Theorem~\ref{hereditary-theorem}
(the detailed proof is deferred to Appendix)]
Assume that $A$ is a $\eps$-strong $\delta,\eps,D$-minimal
statistic for $x$, where $D$ satisfies Lemma~\ref{MSSOmega}.
By Lemma~\ref{MSSOmega} $A$ is informational equivalent 
to $\Omega_{C(A)}$.
We need to prove that  the profile of $[A]$ is close to 
the strong profile of $[A]$.

Let $\mathcal{A}$ be a simple partition and $A_1$ a model from
$\mathcal{A}$ which exists by Lemma \ref{part} applied to $A,x$.
As the total conditional complexities $CT(A_1 \cnd A)$ and  $CT(A \cnd A_1)$
are small, the profiles 
of $A$ and $A_1$ are close to each other. This
also applies to strong profiles. Therefore it suffices to
show that  (the code of) $A_1$ is normal.

Let $(a,b) \in P_{[A_1]}$. The parameters (complexity and log-cardinality)
of $A_1$ are not larger than those of $A$ 
and hence $A_1$ is a sufficient statistic for $x$. 
By Theorem \ref{thprg} we have  
$(a, b + \log|A_1|) \in P_x$ (see Fig. \ref{f6}).
\begin{figure}[h]
\begin{center}
\includegraphics[scale=1]{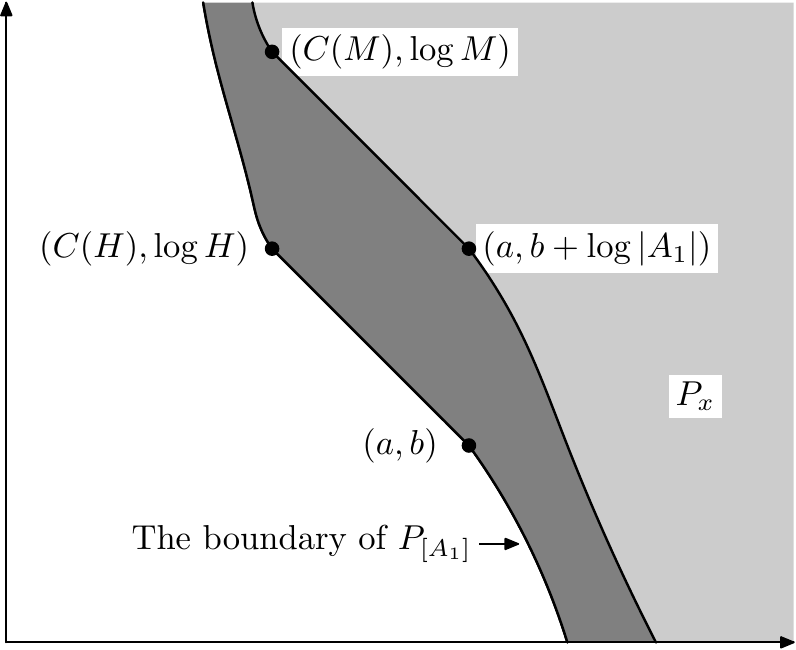}
\end{center}
\caption{$P_x$ is located $\log|A_1|$ higher than $P_{[A_1]}$}
\label{f6}
\end{figure}

As $x$ is normal, the pair  $(a, b + \log|A_1|)$ belongs
to the strong profile of $x$ as well. 
By Lemma \ref{lch} there is a \textbf{strong} model  
$M$ for $x$ 
that has low complexity conditional to $\Omega_{C(M)}$
and whose parameters (complexity, optimality 
deficiency) are not worse 
than those of $A_1$.

We claim that $C(M \cnd A_1)$ is small. As $A$ is informational equivalent 
to $\Omega_{C(A)}$, so is $A_1$. From $\Omega_{C(A)}$ we can compute 
$\Omega_{C(M)}$ (Lemma~\ref{l7}) and then compute $M$ 
(as $C(M \cnd \Omega_{C(M)})\approx0$). This implies that 
$C(M \cnd A_1)\approx0$

However we will need a stronger inequality $CT(M \cnd A_1)\approx0$.
To  find such $M$, 
we apply Lemma~\ref{part} to $M,x$ and change 
it to a model $M_1$ with the same parameters that belongs to a
simple partition $\mathcal M$. Item (2) of Lemma~\ref{part} guarantees
that $M_1$ is also simple given $A_1$ and that $M_1$ is
a strong model for $x$. Since $C(M|A_1)\approx0$, we have 
$C(M_1\cnd A_1)\approx0$ as well.

As $A_1$ lies on the border line of $P_x$ and $C(M_1 \cnd A_1)\approx0$,
the intersection $A_1\cap M_1$ cannot be much less than $A_1$,
that is, $\log|A_1\cap M_1 \cnd \approx\log|A_1|$ (otherwise the model
$A_1\cap M_1$ for $x$ would have much smaller cardinality and 
almost the same complexity as $A_1$). The model $M_1$ can be computed
by a total program from $A_1$ and its index among all $M'\in\mathcal M$ with
$\log|A_1\cap M'|\approx\log|A_1|$. As $\mathcal M$ is a partition,
there are few such sets $M'$. Hence  $CT(M_1 \cnd A_1)\approx 0$.

 
Finally, let
$H=\{A'\in \mathcal{A}\mid \log|A' \cap M_1|=\log|A_1 \cap M_1|\}$.
The model $H$ for $A_1$ is strong because the partition $\mathcal A$
is simple and $CT(M_1 \cnd A_1)\approx 0$.
The model $H$ can be computed from $M_1$, $\mathcal A$ and $\log|A_1 \cap M_1|$.
As  $\mathcal{A}$ is simple, we conclude that
$C(H) \lesssim C(M_1)$.
Finally $\log|H| \le \log|M_1| - \log|A_1|$, because $\mathcal A$
is a partition and thus it has few sets   
that have $\log|A_1 \cap M_1|\approx\log|A_1|$ common elements with $M_1$. 

Thus the complexity of $H$ is not 
larger than that of $M_1$ and the sum of complexity and cardinality
of $H$ is at most $a+b-\log|A_1|$. As the strong profile of $x$ 
has the third property from Theorem~\ref{th1},
we can conclude  that it includes the point $(a,b)$.
\end{proof}

\section*{Acknowledgments}

The author is grateful to  N. K. Vereshchagin
for  advice, remarks and useful discussions.

\section*{Appendix}

\begin{proof}[Proof of Lemma~\ref{pan}]
Let $x$ be the lexicographic first 
string of length $n$ that is not covered by any set 
$A$ of cardinality $2^{n-k}$ and complexity less than $k$.
By a direct counting such a string exists. 
The string $x$ can be computed from $k,n$ 
and the number of halting programs of length
less than $k$ hence $C(x)\le k+O(\log n)$.
To prove that $x$ is normal it is enough  to show that for every $i\le k$ 
there is a $O(\log n)$-strong statistics  $A_i$ 
for $x$ with $C(A_i)\le i+O(\log n)$ and
$\log|A_i| = n - i$.

Let  $A_k=\{x\}$ and 
for $i < k$ let  $A_i$ be the set of all strings of length $n$ whose
the first $i$ bits are the same as those of $x$. 
By the construction $C(A_i) \le i + O(\log n)$ and 
$\log|A_i| = n - i$. 
\end{proof}

\begin{proof}[Proof of Lemma \ref{MSSOmega}]
By Lemma ~\ref{l8} there is $S_{k,m} \ni x$ such that:
\begin{equation}
\label{Skm}
C(S_{k,m} \cnd A) = O(\log n)\text{ and } \delta(x \cnd S_{k,m}) \le \delta(x \cnd A) + O(\log n).
\end{equation}
 From $\delta(x \cnd S_{k,m}) \le \delta(x \cnd A) + O(\log n)$ it follows that $S_{k,m}$ is an $\epsilon + O(\log n)$-sufficient statistic for $x$. 
If the constant $D$ is chosen appropriately, 
then  $S_{k,m}$ is an $\epsilon + D \cdot \log n$-sufficient statistic for $x$, hence, by definition of MSS: 
\begin{equation}
\label{estcs}
C(S_{k,m}) > C(A) - \delta.
\end{equation}
We can estimate $C(\Omega_{C(A)} \cnd A)$ as follows:
\begin{equation}
\label{en3terms}
C(\Omega_{C(A)} \cnd A) \le 
 C(\Omega_{C(A)}\cnd \Omega_{C(S_{k,m})}) + C(\Omega_{C(S_{k,m})} \cnd S_{k,m}) + C(S_{k,m} \cnd A). 
\end{equation}
To prove the lemma it remains to show that every  term of the 
right hand side of this inequality is  $O(\delta + \log n)$. For the third term it follows from (\ref{Skm}).

To prove it for the first term note that $|C(A) - C(S_{k,m}) \cnd \le \delta + O(\log n)$ by (\ref{Skm}) and (\ref{estcs}). Now the inequality $C(\Omega_{C(A)}\cnd \Omega_{C(S_{k,m})}) \le \delta + O(\log n)$ follows from the following simple lemma.

\begin{lemma}
\label{ome}
Let $a, b$ be some integers. Then 

$C(\Omega_a \cnd \Omega_b) \le |a-b| + O(\log (a+b))$.
\end{lemma}
\begin{proof}

Consider two cases. If $b \ge a$, then $C(\Omega_a \cnd \Omega_b) = O(\log b)$ by the first statement of  Lemma \ref{l7}.

If $b < a$ we get $C(\Omega_b \cnd \Omega_a) = O(\log a)$ by the same argument. By symmetry of information we get:
$$C(\Omega_a \cnd \Omega_b) = C(\Omega_a) - C(\Omega_b) + O(\log (a+b)).$$
To conclude the required statement it remains to recall that $C(\Omega_a) = a + O(\log a)$ and $C(\Omega_b) = b + O(\log b)$ by Lemma \ref{l7}.
\end{proof}

Now it remains to show that the second term $C(\Omega_{C(S_{k,m})} \cnd S_{k,m})$
of the right hand side of the inequality (\ref{en3terms}) is $O(\log n)$. 
This is an easy corollary from the second item of Lemma~\ref{l7}
and the equality $C(S_{k,m})=k-m+O(\log k)$. 
\end{proof}

\begin{proof}[Proof of Lemma \ref{lch}]
Let $E$ be a statistic for $x$. Denote by $f(E)$ a statistic for $x$ 
that is not worse than $E$ and is equivalent to $\Omega_t$ 
for some $t$, that is, a statistic that exists by 
Theorem~\ref{l8}: 
\begin{align*}
C(f(E) \cnd E) = O(\log n),\quad 
\delta(x \cnd f(E)) \le \delta(x \cnd E) + O(\log n),\\
C(f(E) \cnd \Omega_{C(f(E))}) = O(\log n),\quad
C(\Omega_{C(f(E))} \cnd f(E)) = O(\log n). 
\end{align*}
Denote by $g(E)$ a statistic for $x$ such that 
$$
C(g(E)) < C(E) + \alpha,\quad \log|g(E)| < C(E) + \alpha,
\quad g(E) \text{ is $\epsilon$-strong.}
$$ 
Such 
model $g(E)$  exists for every $E$ because $x$ is $\epsilon,\alpha$-normal. 

Consider the following sequence: 
$$
A_1= A, \quad B_1 = f(A_1), \quad A_2 = g(B_1), \quad B_2 = f(A_2), \dots
$$

Let us call a pair $A_iB_i$ a \emph{big step} if $C(A_i) - C(B_i) > \sqrt{n}$. Denote by $N$ the minimal integer such that $A_N B_N$ is not a big step. Let us show that $N = O(\sqrt{n})$. Indeed, $C(A_{i+1}) < C(B_i) + \alpha$ and thus 
$C(A_{i+1}) - C(A_i) > \sqrt{n} - \alpha$ for every $i < N$. 
On the other hand $C(A_1) \le C(x) + CT(A_1 \cnd x) \le n + \epsilon$. 
Therefore $N \cdot (\sqrt{n} - \alpha) \le n + \epsilon$. Since 
$\alpha < \sqrt{n}/2$, $\epsilon \le n$ we have 
$N = O(\sqrt{n})$. 

Let $H=A_N$. Let us show that $H$ satisfies all the requirements.

1) $A_N$ is an $\epsilon$-strong model for $x$ by definition of $g$. 
 
2) Let us estimate of $\delta(x \cnd A_N)$. 
We have $\delta(x \cnd A_{i+1}) \le \delta(x \cnd B_i) + 2\cdot \alpha$ and 
$\delta(x \cnd B_i) \le \delta(x \cnd A_i) + O(\log n)$ for every $i$. So 
$$
\delta(x \cnd A_N) \le \delta(x \cnd A_1) + N\cdot(2\alpha + O(\log n)) \le \delta(x \cnd A_1) + O(\alpha + \log n)\cdot \sqrt{n}.
$$ 

In a similar way we can estimate the complexity of $A_N$:
$C(B_i) < C(A_i) - \sqrt{n}$ if $i < N$ and $C(A_{i+1}) < C(B_i) + \alpha$. 
As $\alpha < \sqrt{n}/2$ we conclude that 
$C(A_{i+1}) < C(A_i)$ for $i < N$.  Hence $C(A_N) \le C(A_1)$.

3) To estimate $C(B_N \cnd A_N)$ we use the following inequality:
$$C(A_N \cnd \Omega_{C(A_N)}) 
\le C(A_N \cnd B_N)+ 
  C(B_N \cnd \Omega_{C(B_N)}) + C(\Omega_{C(B_N)} \cnd \Omega_{C(A_N)}).$$

It remains to show that all terms in the right hand side 
are equal to $O(\sqrt{n})$.
This bound holds for the first term because, as $A_N B_N$ is not a big step. 
The second term is equal to $O(\sqrt{n})$ by the definition 
of $f$. For the third term we use 
the inequalities $|C(A_N) - C(B_N)| < \sqrt{n}$ (the pair $A_N,B_N$ 
is not a big step and $C(B_N \cnd A_N) = O(\log n)$) and Lemma~\ref{ome}.
\end{proof}

\begin{proof}[Detailed proof of Theorem\ref{hereditary-theorem}]
Again we will use the notations from the sketch of proof.

\emph{Step 1: From  $A$ to $A_1$.}

As the model $A$ is $\epsilon$-strong for $x$, 
by Lemma ~\ref{part} there is an $\epsilon + O(\log n)$-strong statistic  $A_1$
for 
$x$ that belongs to an $\epsilon + O(\log n)$-simple partition $\mathcal{A}$ such that
\begin{equation}
\label{total1}
CT(A \cnd A_1) < \epsilon + O(\log n)
\end{equation}
\begin{equation}
\label{total2}
CT(A_1 \cnd A) < \epsilon + O(\log n)
\end{equation}
\begin{equation}
\label{size}
|A_1| \le |A|.
\end{equation}
We will show that $P_{[A_1]}$ is close to $P_{[A_1]}^{O((\epsilon + \delta +  \log n)\cdot \sqrt{n})}$
and then we will prove a similar statement for $A$.

Let $(a,b) \in P_{[A]}$. We need to show that $(a,b)$ is close  to 
$P_{[A_1]}^{O((\epsilon + \delta +  \log n)\cdot \sqrt{n})}$. This is 
straightforward, if $a \ge C(A)$.
Therefore  we will assume that $a < C(A)$.  
From ~(\ref{total2})  it is easy to see that 
$(a + O(\epsilon + \log n), b + O(\epsilon + \log n)) \in P_{[A_1]}$.

The set $A$ is an $\epsilon$-sufficient statistic for $x$. From this,  ~(\ref{total2}) and ~(\ref{size})  it follows that $A_1$ is an $O(\epsilon + \log n)$-sufficient statistic for $x$. 

\emph{Step 2: From $A$ to $M$.}

From now on we will omit terms of the order 
$O((\epsilon + \delta +  \log n)\cdot \sqrt{n})$.

By Theorem ~\ref{thprg} from sufficiency of $A_1$ it follows that $(a , b + \log|A_1|) \in P_x$. As $x$ is $\eps,\eps$-normal, 
a point with similar parameters belongs to $P_x^{\epsilon}$. A statistic with corresponding parameters can be improved by Lemma~\ref{lch}, i. e., 
there is an $\epsilon$-strong statistic $M$ for $x$ such that:
\begin{equation}
\label{m111}
C(M \cnd \Omega_{C(M)}) = 0 \text{, }C(M) \le a \text{,}
\end{equation}
\begin{equation}
\label{m113}
\text{and }\delta(x \cnd M) \le a + b + \log|A_1| - C(x). 
\end{equation}

\emph{Step 3: From $M$ to $M_1$.}

By Lemma ~\ref{part} we can transform $M$ to an $\epsilon + O(\log n)$-strong statistic $M_1$ for $x$ that belongs to an $\epsilon + O(\log n)$-simple 
partition $\mathcal{M}$ and whose parameters are not worse than
those of $M$:
\begin{eqnarray}
\label{M_1}
CT(M \cnd M_1) = 0, \quad
CT(M_1 \cnd M) = 0, \quad
|M_1| \le |M|,\\ \label{omega}
C(M_1 \cnd \Omega_{C(M_1)}) = 0,\\
\label{complexity}
C(M_1) \le a,\\ 
\label{optim}
\delta(x \cnd M_1) \le a + b + \log|A_1| - C(x). 
\end{eqnarray} 

Now we need the following
\begin{lemma}
$\log|A_1\cap M_1| = \log|A_1|$
(up to  $O((\epsilon + \delta +  \log n)\cdot \sqrt{n})$).
\end{lemma}
\begin{proof}[Proof of Lemma]
The model 
$A$ is a $\delta,\eps,D$-MSS for $x$, 
hence by Lemma ~\ref{MSSOmega}  $C(\Omega_{C(A)} \cnd A) = 0$. 
On the other hand we have $C(A \cnd A_1)= 0$. Hence
\begin{equation}
\label{omegaA}
C(\Omega_{C(A)} \cnd A_1) = 0.
\end{equation}
Recall  that we assume that 
$a < C(A)$. Inequality (\ref{complexity}) states that $C(M_1) < a$ and therefore
$C(M_1) < C(A)$. 
Hence, from  Lemma ~\ref{l7}  it follows that 
$$
C(\Omega_{C(M_1)} | \Omega_{C(A)}) = 0.
$$
 From this, (\ref{omega}) and (\ref{omegaA}) it follows that $C(M_1 \cnd A_1) = 0$. Obviously, we have 
 $C(M_1\cap A_1) \le  C(A_1) + C(M_1 \cnd A_1) $ and thus 
\begin{equation}
\label{cap}
C(M_1\cap A_1) \le C(A_1). 
\end{equation}
As  $A_1$ is  a sufficient statistic for $x$ we conclude that 
$$
\log|A_1\cap M_1|  + C(M_1\cap A_1) \ge  C(A_1) + \log|A_1|.
$$ 
From this and (\ref{cap}) it follows that $\log|A_1\cap M_1| \ge \log|A_1|$.
\end{proof}

\emph{Step 4: Constructing $H$.}

Denote by $H$ the family of sets from $\mathcal{A}$ which have the
same size of intersection with $M_1$ as $A_1$ up to a factor of 2:
$$
H = \{A' \in \mathcal{A}| \lfloor \log|A' \cap M_1| \rfloor  = \lfloor \log A_1 \cap M_1 \rfloor \}.
$$

As $\mathcal{A}$ is partition, we have  
$|H| \le |M_1|/(2 \cdot |A_1\cap M_1|)$. Therefore we have 
$$\log|H| \le \log|M_1| - \log|A_1|. $$
We can compute $H$ 
from $M_1$, $\mathcal{A}$ and $\lfloor \log A_1 \cap M_1 \rfloor$, so:
$$
C(H) \le C(M_1) + C(\mathcal{A}) = C(M_1). 
$$
By a similar reason we have 
$$
CT(H \cnd A_1) \le CT(M_1 \cnd A_1) + C(\mathcal{A}) + O(1) \le CT(M_1 \cnd A_1).
$$

To estimate $CT(M_1 \cnd A_1)$ recall that $\mathcal{M}$ is a partition, 
so there are at most $|A_1|  / (2 \cdot |A_1 \cap M_1|)$ elements from $\mathcal{M}$ who have $ |M_1 \cap A_1|/2$ common strings with $A_1$. Thus we have: 
 $$CT(H \cnd A_1) \le CT(M_1 \cnd A_1) \le \log|A_1| - \log|A_1\cap M_1| + 1 + C(\mathcal{M}) \approx 0.$$
Thus $H$ is strong statistic for $A_1$ of complexity at most $C(M_1)$ and log-size at most $\log|M_1| - \log|A_1|$: 
\begin{equation}\label{parH}
(C(M_1) ,\log|M_1| - \log|A_1|) \in P_{[A_1]}^{O(\sqrt{n} + \epsilon + \delta)}.
\end{equation}

\emph{Step 4: Back to $A$}.

Inequality (\ref{complexity}) states that $a \ge C(M_1)$.  
Since a strong profile also has the third property from Theorem~\ref{th1},
we can add  $a - C(M_1)$ to the first component 
and subtract it from the second component of 
the left hand side of~\eqref{parH} 
(i. e. make the statistic smaller but more complex): 
$$
(a ,\log|M_1| - \log|A_1| - a + C(M_1) ) \in P_{[A_1]}^{O(\sqrt{n} + \epsilon + \delta)}.$$
By ~(\ref{optim}) the second component becomes less than $b$,   i.e.
$(a, b) \in P_{[A_1]}^{O(\sqrt{n} + \epsilon + \delta)}$.

We have shown that there is a set $B \ni [A_1]$  such that:
$$
C(B) = a ,\quad \log |B| = b,\quad CT(B \cnd [A_1]) = 
O(\sqrt{n} + \epsilon + \delta).
$$  

Equation (\ref{total1}) states that
there is a total programs $p$  of length $\epsilon + O(\log n)$ such that  $p([A_1]) = [A]$.
Consider the set  $D:= \{ p(t) \cnd t \in B\}$. 
The set $D$ is the required model for $A$.
Indeed, we have 
$[A] \in D$, $\log|D| \le \log|B|$, $C(D) \le C(B)  + l(p) + O(1) = a + O(\log n + \epsilon)$, $CT(D \cnd [A]) \le CT(D \cnd B) + CT(B \cnd A_1) + CT(A_1 \cnd A) \le O(\sqrt{n} + \epsilon + \delta)$. Therefore we have 
$$ 
(a + O(\log n + \epsilon), b+ O((\epsilon + \delta +   \log n)\cdot \sqrt{n}) ) \in P_{[A]}^{O(\sqrt{n} + \epsilon + \delta)}.
$$
\end{proof}

\end{document}